\documentclass[trackchanges,twocolumn,times]{aastex63}

\shorttitle{AR  upflows}                 

\shortauthors{Brooks et al. }
\journalinfo{To be submitted to the Astrophysical Journal}

\usepackage{grffile}

\begin{document}


\title{The Formation and Lifetime of Outflows in a Solar Active Region}

\author[0000-0002-2189-9313]{David H.\ Brooks}
\affil{College of Science, George Mason University, 4400 University Drive, Fairfax, VA 22030, USA}
\affil{Hinode Team, ISAS/JAXA, 3-1-1 Yoshinodai, Chuo-ku, Sagamihara, Kanagawa 252-5210, Japan}

\author[0000-0001-9457-6200]{Louise Harra}
\affil{PMOD/WRC, Dorfstrasse 33 CH-7260 Davos Dorf, Switzerland}
\affil{ETH-Zurich, H\"onggerberg campus, HIT building, Z\"urich, Switzerland}

\author[0000-0002-1989-3596]{Stuart D. Bale}
\affil{Physics Department and Space Sciences Laboratory, University of California, Berkeley, CA 94720-7450, USA}

\author[0000-0001-7090-6180]{Krzysztof Barczynski}
\affil{PMOD/WRC, Dorfstrasse 33 CH-7260 Davos Dorf, Switzerland}
\affil{ETH-Zurich, H\"onggerberg campus, HIT building, Z\"urich, Switzerland}

\author[0000-0001-9311-678X]{Cristina Mandrini}
\affil{Instituto de Astronom\'ia y F\'isica del Espacio (IAFE), CONICET-UBA, Buenos Aires, Argentina }

\author[0000-0002-4980-7126]{Vanessa Polito}
\affil{Bay Area Environmental Research Institute, NASA Research Park,  Moffett Field, CA 94035, USA}
\affil{Lockheed Martin Solar \& Astrophysics Laboratory,
      Org. A021S, Bldg. 252, 3251 Hanover St., Palo Alto, CA 94304, USA}

\author[0000-0001-6102-6851]{Harry P. Warren}
\affil{Space Science Division, Naval Research Laboratory, Washington, DC 20375, USA}


\begin{abstract}
Active regions are thought to be one contributor to the slow solar wind. Upflows in EUV coronal spectral lines are routinely osberved at their boundaries, and provide the most direct way for upflowing material to escape into the heliosphere. The mechanisms that form and drive these upflows, however, remain to be fully characterised. It is unclear how quickly they form, or how long they exist during their lifetimes. They could be initiated low in the atmosphere during magnetic flux emergence, or as a response to processes occuring high in the corona when the active region is fully developed. On 2019, March 31, a simple bipolar active region (AR 12737) emerged and upflows developed on each side. We used observations from Hinode, SDO, IRIS, and Parker Solar Probe (PSP) to investigate the formation and development of the upflows from the eastern side. We used the spectroscopic data to detect the upflow, and then used the imaging data to try to trace its signature back to earlier in the active region emergence phase. We find that the upflow forms quickly, low down in the atmosphere, and that its initiation appears associated with a small field-opening eruption and the onset of a radio noise storm detected by PSP. We also confirmed that the upflows existed for the vast majority of the time the active region was observed. These results suggest that the contribution to the solar wind occurs even when the region is small, and continues for most of its lifetime.
\end{abstract}



\section{Introduction}

Outflows from the edges of active regions have become a focus of many studies since they were noted in early observations
from Hinode \citep{Kosugi2007}. These outflows are of great interest because of their likely contribution to the slow solar wind
\citep{Sakao2007,Harra2008,Doschek2008}. Upflows in active regions are most obvious in hot spectral lines formed around 1--2\,MK \citep{DelZanna2008,Warren2011},
and occur in dark areas where the line intensities are faint, especially at the active region edges \citep{Doschek2008}.
They show bulk plasma motions on the order of tens of km s$^{-1}$ \citep{DelZanna2008}. A higher speed component, reaching hundreds of km s$^{-1}$,
is also often present in the blue wing of EUV spectral lines \citep{Bryans2010,Peter2010,Tian2011,Brooks2012}. Plasma composition measurements
and simple mass flux estimates have strengthened the idea of a connection to the slow solar wind \citep{Brooks2011,Brooks2015,Brooks2021}. Indeed different
composition signatures in the outflows may help explain variability in the slow wind \citep{Brooks2020}, and the evolution of the  upflows
has even been linked to radio noise storms detected close-in to the Sun by Parker Solar Probe \citep{Harra2021}. 

Linking remote sensing
and in-situ observations is a key goal of both Parker Solar Probe \citep[PSP,][]{Fox2016} and Solar Orbiter \citep{Muller2020}. Detailed recent reviews focusing
on active region outflows are given by \cite{Hinode2019} and Tian et al. (2021), while an extensive review including their contribution to the
solar wind was presented by \cite{Abbo2016}.

\begin{figure*}
\centering
\includegraphics[width=1.0\textwidth]{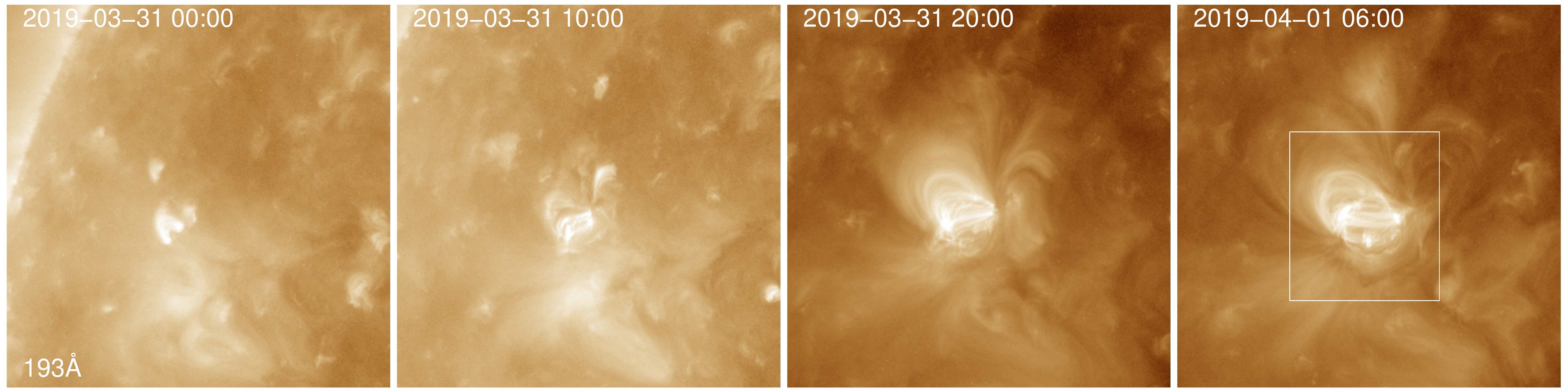}
\caption{ AIA 193\,\AA\, images showing the emergence and growth of AR 12737 from 00\,UT on March 31 to 06\,UT on April 1. The white box shows the FOV
of the IRIS observations in Figure \ref{fig4}. }
\label{fig1}%
\end{figure*}

There remain several outstanding issues with our understanding of the upflows/outflows, and these have significant implications for their contribution to the slow 
solar wind. In particular, it is not clear how early they form in the emergence phase of an active region, 
nor how long they persist during its lifetime. Do upflows become outflows and contribute to the slow wind all the time they exist, or just for a shorter fraction
of the active region lifetime? Clearly active regions are a more significant contributor to the slow wind if 1) the  upflows exist longer, and 2)
even small regions have  upflows.

One of the difficulties in studying the early formation phase is that the spectroscopic instruments used to measure plasma flows around active regions
have small fields-of-view (FOV) and slow slit scanning times. This makes it challenging to catch active regions as, or soon after, they emerge. \cite{Harra2010} 
present observations of emerging flux within an already developed active region that was observed by the EUV Imaging Spectrometer \citep[EIS,][]{Culhane2007}
on Hinode. The interaction between newly emerging and pre-existing opposite polarity magnetic field formed a ring of strong  upflows at the active region
edge. This formed quickly - within 12 hours - but we should note that the magnetic topological conditions were very favourable since large scale  upflows were already
present in the overlying developed active region. These may already have opened the field surrounding the active region. It is unclear if  upflows form
as quickly in an isolated active region.

These ideas are closely connected with the  upflow formation mechanism itself, which is 
another unresolved issue.
Perhaps the most popular picture is that closed field loops in the active region core are opened by interchange reconnection with the open fields in its surroundings.
This can occur at quasi-separatrix layers \citep{Baker2009,Mandrini2015}, where there are strong gradients in magnetic connectivity, and may be driven by the active region
emergence and expansion \citep{Murray2010,DelZanna2011}. This process provides a mechanism to transfer the closed (solar wind-like) composition of the hot
core loops onto open magnetic fields. 

Again, a key question is when and where? If emergence is the driver, this could happen quickly and low down in the atmosphere.
Conversely, it could be that even after  upflow formation, the parent active region needs to expand and interact
with high lying magnetic field before the  upflow opens to the heliosphere. 
Some studies have shown that the  upflow magnetic field is not always open \citep{Edwards2016}, while others suggest that some fraction
of the  upflow mass flux also flows through connections to distant active regions \cite{Boutry2012}. 

Another possibility is the direct injection of mass and energy into the  upflows by chromospheric jets \citep{DePontieu2009}.
Recently \cite{Polito2020} found signatures of the  upflows in chromospheric spectral lines observed by the Interface Region Imaging Spectrometer 
\citep[IRIS,][]{DePontieu2014}. Several chromospheric and transition region lines showed different behavior in the  upflows 
compared to the cores of two active regions.
It is still unclear, however, if these signatures are evidence of the driving mechanism of the  upflows operating at low heights.
It is also possible that they are revealing a chromospheric response to the changed coronal environment of open-field regions. 
Based on the multi-wavelength analysis of Hinode/EIS and IRIS data, Barczynski et al. (2021) argue that at least three parallel 
mechanisms generate the plasma upflow, and that these mechanisms are localized in the chromosphere, transition region, and corona.

AR 12737 emerged on the Earth facing solar disk on 2019, March 31, during the 2nd PSP encounter, and was targetted 1--2 days later by EIS and IRIS.
From a case study of AR 12737, we will show evidence that
1) the  upflows in this region are formed low down, in the early emergence phase, when it is still relatively small, and 2) once formed, 
the  upflows exist for the entire observed lifetime of the region.

\section{Observations} \label{sec:obs}

In this work we use several datasets from the Atmospheric Imaging Assembly \citep[AIA,][]{Lemen2012}
on board the Solar Dynamics Observatory \citep[SDO,][]{Pesnell2012}. These were downloaded via the web-based interface to the
Joint Science Operations Center (JSOC) at Stanford, are calibrated and correspond to level 1.5. 
We retrieved 193\,\AA\, data for two long duration time-periods at 156\,s and 600\,s cadence, and a high cadence (12\,s)
multi-wavelength (304\,\AA, 171\,\AA, 193\,\AA, and 211\,\AA) dataset around the time of the  upflow onset.

For Doppler velocity maps of the active region corona we use EIS measurements.
EIS is a dual spectrograph that observes in the 171--211\,\AA\, and 245--291\,\AA\, wavelength ranges with a 
spectral resolution of 22.3\,m\AA. To account for instrumental effects (defective pixels, dark current, cosmic ray hits) we reduced
the data using the eis\_prep SolarSoftware routine. The observations we analyze are 261$''\times512''$ field-of-view (FOV) rasters of AR 12737
using the 2$''$ slit with coarse scan steps of 3$''$. The exposure time was 40\,s.

To obtain the Doppler velocity maps we fit the strong Fe XII 195.119\,\AA\, spectral line. We used a double Gaussian function to take account of 
the density sensitive Fe XII blend at 195.179\,\AA. Measured line centroids were converted to velocities after first correcting for spectral
motion across the CCD, due to instrument structure temperature variations around the Hinode orbit, using the artificial neural network
model of \cite{Kamio2010}. We then calibrated the velocities to a reference wavelength obtained by averaging the top part of the FOV, and 
finally removed a residual orbital variation that was present following the strategy of \cite{Brooks2020}. This last step was necessary
since the neural network model uses data from early in the mission that are less applicable to recent observations.
The accuracy is on the order of 4--5\,km s$^{-1}$, though we do not use any actual values in this study.

\begin{figure*}
\centering
\includegraphics[viewport = 70 20 1566 492,width=1.0\textwidth]{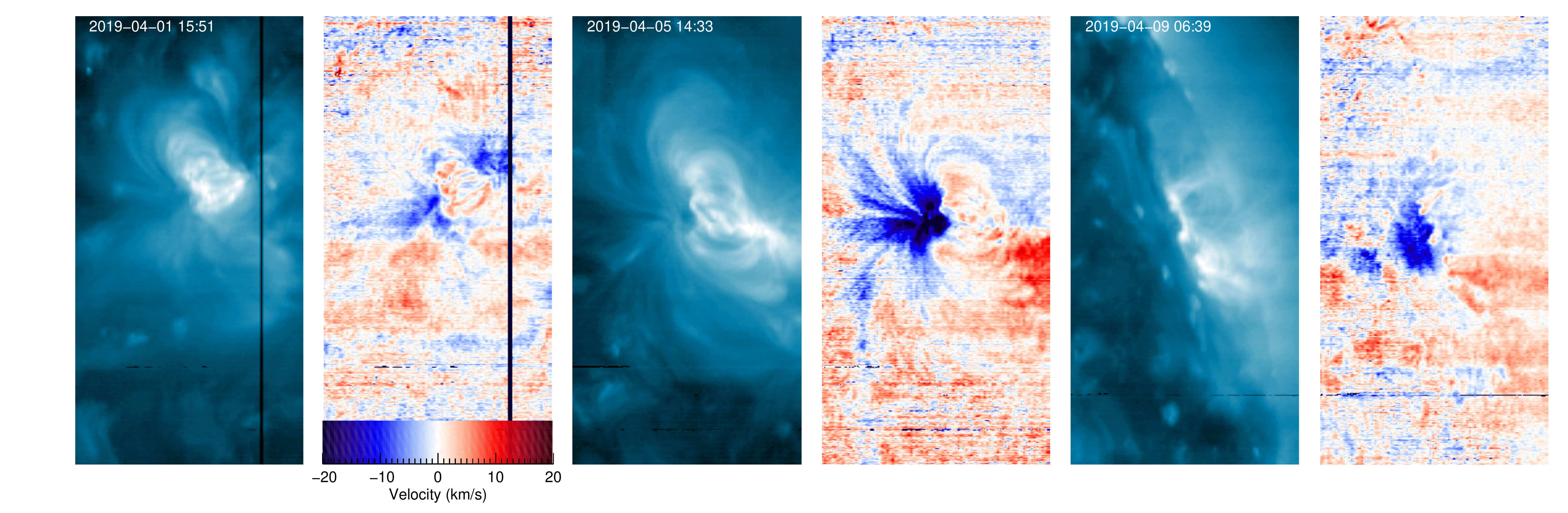}
\caption{ EIS intensity and Doppler velocity maps of AR 12737 between April 1 and 9. The maps were derived from the
Fe {\sc xii} 195.119\,\AA\, spectral line.
The colour bar shows the velocity range. }
\label{fig2}%
\end{figure*}

\begin{figure}[h]
\centering
\includegraphics[viewport = 0 190 360 550,clip=true,width=0.50\textwidth]{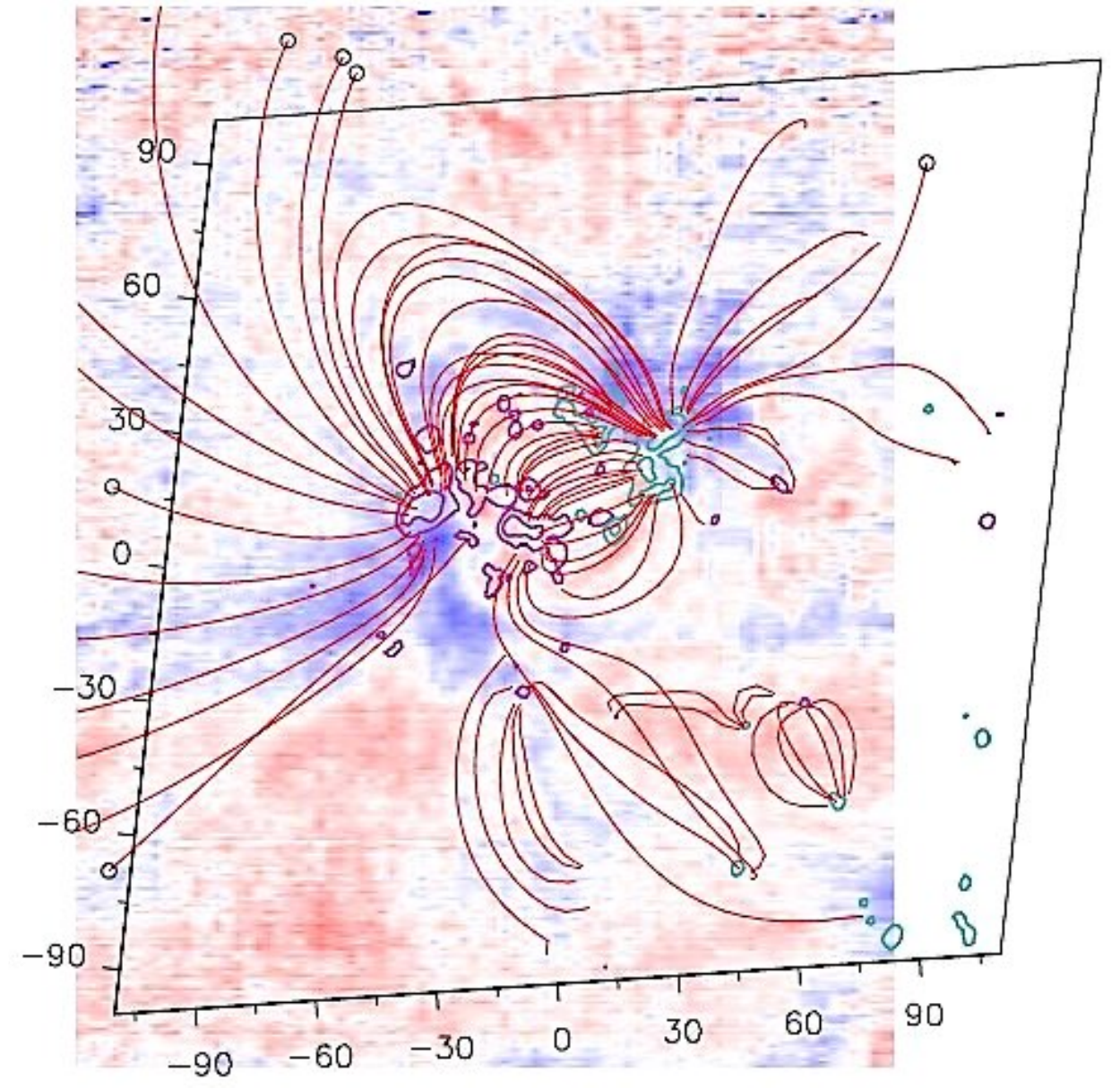}
\caption{ Magnetic field model for AR 12737 on April 1. The extrapolated field lines are overlaid in red on the EIS Doppler velocity map obtained at
15:51\,UT\, and already shown in Figure \ref{fig2}. The axes show the model scale in Mm with the origin set at the active region 
center. The magenta and cyan contours show the line-of-sight magnetic field for positive and negative values of $\pm$50--500\,G.
Field lines ending with a circle leave the computational box and are potentially open. }
\label{fig3}
\end{figure}

For observations of chromospheric structure and velocities we used IRIS. The IRIS instrument observes two wavelength bands in the far and
near-ultraviolet (FUV \& NUV) covering 1332--1407\,\AA\, and 2783--2835\,\AA. IRIS has a slit-jaw imager and spectrograph. Here we focus on 
observations in the Si IV 1393\,\AA, C II 1335\,\AA, and Mg II 2796\,\AA\, spectral lines. These cover the transition region, upper chromosphere, 
and mid to upper chromosphere, respectively. We use level-2 data, and these are processed to account for instrumental effects (dark current, 
geometric effects, flat-field, orbital wavelength variation). The observations we analyze are 129$''\times126''$ FOV rasters 
constructed from coarse (2$''$ step scans) at a spatial resolution of 0.33--0.4$''$. The exposure time was 15\,s.

Doppler velocity maps of the upper chromosphere and transition region were derived from spectral fits to the C II 1335\,\AA\, and Si IV 1393\,\AA\, 
lines. The Si IV 1393\,\AA\, velocities are derived from the line peak of a single Gaussian fit.
The C II 1335\,\AA\, line profiles are complex and can be singly or doubly peaked. The velocities here are derived from the peak if the profile 
has a single peak, but are computed from the line reversal position if
the profile is double peaked. We used the algorithm of \cite{Rathore2015} to identify the profile peaks. 
We also constructed a map of the Mg II 2796\,\AA\, $k2$ asymmetry i.e. the difference in intensity between the peaks
in the red and blue wings of the line profile. A positive asymmetry can imply upflows due to increased absorption in the blue wing, while a negative
asymmetry can imply downflows due to increased absorption in the red wing. 
We analyzed the Mg II spectral profiles using the iris\_get\_mg\_features\_lev2 procedure available in the IRIS branch of SolarSoftware \citep{Freeland1998}.

We also compare our analysis of AIA images with type III radio noise data recorded by the FIELDS \citep{Bale2016} Radio Frequency Spectrometer 
\citep[RFS,][]{Pulupa2017} on PSP. The RFS obtains full Stokes parameters using low- and high-frequency receivers covering a wide range
from 10.5\,kHz to 19.17\,MHz. The usual spectral cadence is 7\,s.
We only use a subset of the RFS encounter 2 data, described in detail by \cite{Harra2021}, for illustration. We focus only on the frequency 
of maximum Stokes intensity. 
The data were reduced as described in \cite{Harra2021}.

\begin{figure*}
\centering
\includegraphics[viewport = 25 30 708 170,clip=true,width=1.0\textwidth]{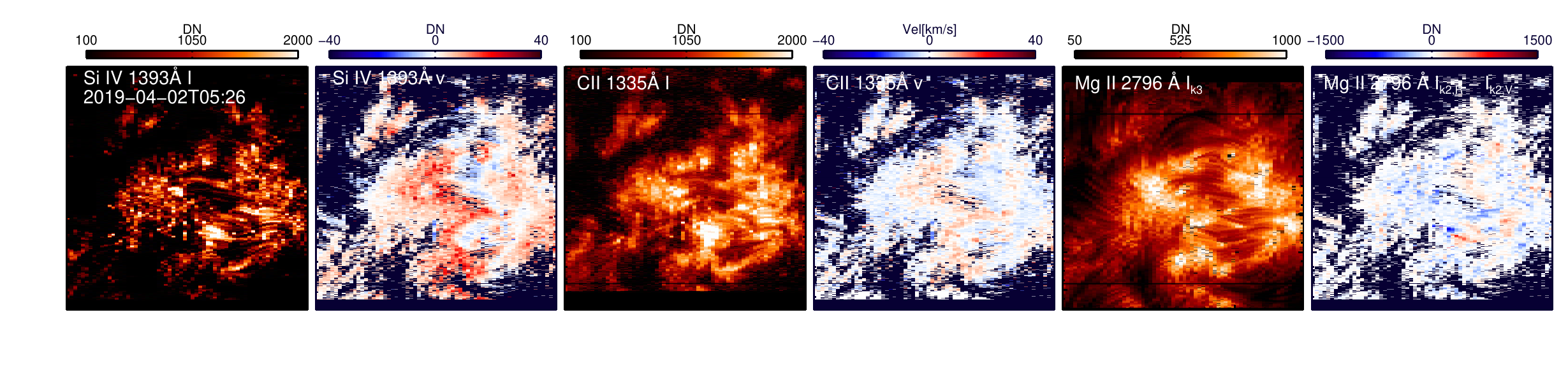}
\caption{ IRIS Si IV 1393\,\AA\, and C II 1335\,\AA\, intensity and velocity maps AR 12737 on April 2. The last two panels show the
Mg II 2796\,\AA\, $k3$ intensity and asymmetry between the $k2_R$ and $k2_v$ peaks. The FOV is shown by the white box in Figure \ref{fig1}. 
}
\label{fig4}%
\end{figure*}

\section{Analysis results} \label{sec:results} 

\subsection{Evolution of the  upflows} \label{sec:results1}

AR 12737 was a quiescent active region. Despite its emergence the GOES X-ray flux remained below B-class while
it was the sole region on disk. Figure \ref{fig1} shows AIA 193\,\AA\, images to give an overview of its emergence phase.
It appears to have evolved from a cusp-shaped loop arcade around 00\,UT on March 31. By 10\,UT it is clear that a new
active region is forming, and this development phase lasts several hours. By 20\,UT the typical structure of an active region
has formed. 
In Figure \ref{fig1} we can see 
a hot core loop arcade, high lying million degree loops, and dark channels from the active region edge, which often show
propagating motions in imaging data. These dark areas at the active region edge typically show upflow signatures when observed by EIS \citep{Doschek2008}. 
In section \ref{sec:results2} we attempt to estimate these time periods more quantitatively.

EIS moved to observe AR 12737 from 15:51\,UT on April 1, and tracked the region across the solar disk, making a final 
slit scan at 06:39\,UT on April 9. Figure \ref{fig2} shows an overview of these observations. EIS spectroscopically confirms the existence
of  upflows in the dark channels on both the east and west sides of AR 12737 as early as 16\,UT on April 1. 

\cite{Harra2021} used linear force-free field models to establish the global coronal structure of AR 12737, and discussed the expansion and development of the eastern upflow after April 1 in detail. They show that the area of the blue-shifted
 upflow expands by a factor of 10 between April 1 and April 4, that the region is associated with large scale
magnetic field lines in their model, and that the area associated with these large scale field lines increases as the AR expands. 
This seems to be driven by the expansion of closed loops to the south east of the AR,
which appear as red-shifted in the EIS velocity map of April 1 (Figure \ref{fig2} left panel).
We show an example of their magnetic field model in Figure \ref{fig3}, overlaid on an EIS Doppler velocity map to highlight the associations
between the extrapolated field lines and the active region outflows. Large scale field lines are clearly seen rooted in the positive polarities at
the base of the blue-shifted outflow on the east side (around coordinates [-30,0]). The field lines are a composite from different models and the details 
are described in \cite{Harra2021}.

Figure \ref{fig2} also shows that the eastern  upflow is still clearly visible on April 9. Projection effects, however, mean that we can no longer
detect blue-shifted emission in the western  upflow.

IRIS scanned AR 12737 from 05:17--05:36\,UT\, on April 2 ($\sim$13.5 hours after the first EIS scan). Figure \ref{fig4} shows intensity and
velocity diagnostics from the transition region (Si IV 1393\,\AA) down through the upper (C II 1335\,\AA) and mid- to upper chromosphere (Mg II 2796\,\AA). 
\cite{Polito2020} previously
reported chromospheric signatures of coronal  upflows in IRIS observations of two active regions. They found that the average redshift in the Si IV 1393\,\AA\,
line was reduced, the C II 1335\,\AA\, line was slightly blueshifted, and that Mg II k2 asymmetries were present,
in the  upflows compared to the active region core. Figure \ref{fig4} indicates that these characteristics are present in the eastern  upflow of AR 12737.
Note, for example, the strong redshifts in the core of the region in the Si IV 1393\,\AA\, velocity map, and the reduced redshift (occasional blueshift)
to the east (second panel of Figure \ref{fig4}. 

These observations provide another example of chromospheric signatures of the  upflows, but they 
appear to be reduced in AR 12737 compared to the active regions analyzed by \cite{Polito2020}. Visual inspection of the locations compared to 
the EIS data suggests that the spatial correlation is not as clear, especially as the morphology of the  upflow has already evolved in the 13 hours
between the EIS and IRIS observations. 
The weaker signatures are probably due to the fact that AR 12737 was not yet
fully developed on April 2, so the spectral lines were also weaker, and the  upflow region had not yet expanded to the extent observed on April 4. 
These results do indicate, however, 
that chromospheric signatures of the  upflows have already appeared by two days into the lifetime of the active region. 

\subsection{Formation of the  upflows} \label{sec:results2}
We have attempted to pinpoint the transition from an emerging flux region to a formed active region with  upflows using 
a combination of time-slice intensity tracking, cross-correlation analysis of images, and simple visual
inspection. Without spectroscopic data of the very early emergence phase this is, of course, difficult to confirm, but here we
discuss evidence that the  upflows may have formed as early as 12--16\,UT on March 31.

\begin{figure*}
\centering
\includegraphics[width=1.0\textwidth]{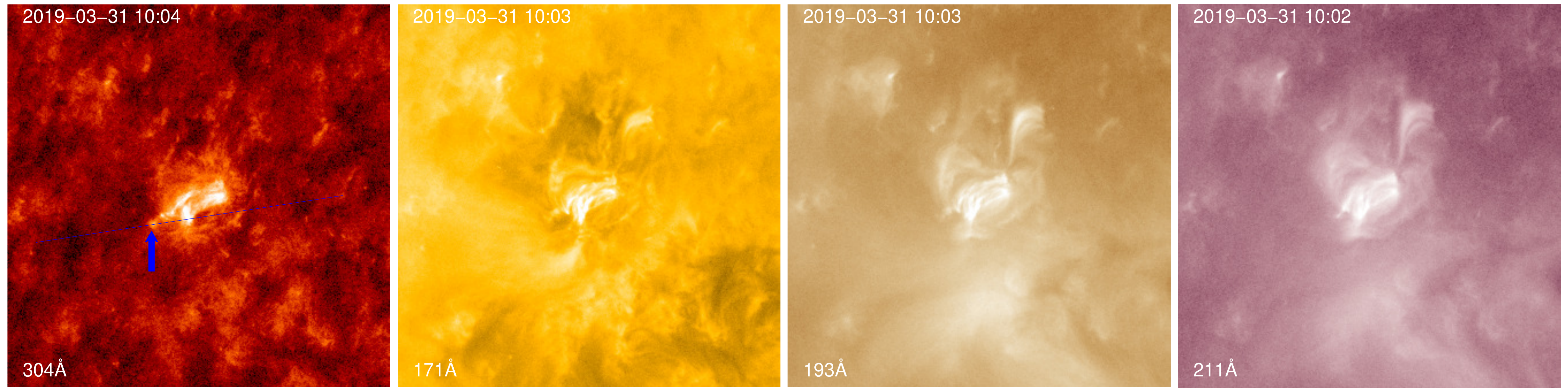}
\caption{ AIA 304\,\AA, 171\,\AA, 193\,\AA, and 211\,\AA\, images showing the small eruption at 10\,UT on March 31.
The blue arrow on the 304\,\AA\, image points out the eruption from the solar east side of AR 12737.
This image is linked to a 30\, min animation in the online version of the manuscript. }
\label{fig5}%
\end{figure*}

\begin{figure}[h]
\centering
\includegraphics[viewport = 100 20 560 512,clip=true,width=0.50\textwidth]{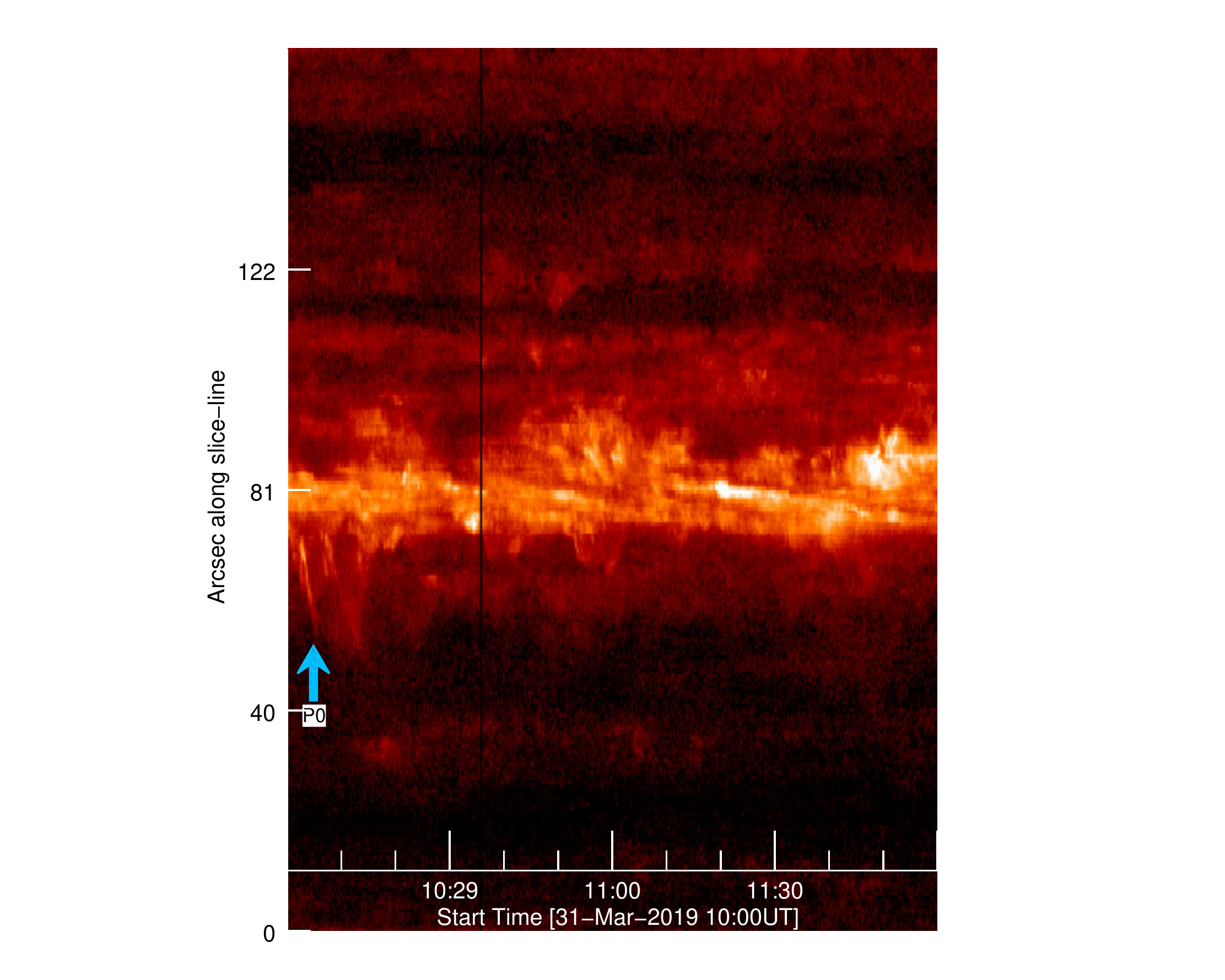}
\caption{ Space-time analysis showing the mini-eruption in the 304\,\AA\, data. The space-time plot corresponds to the thin
blue line in Figure \ref{fig5}. The data were taken at a high cadence of 12\,s between 10--12\,UT on March 31.
The sky blue arrow points out the start of the small eruption discussed in the text.
\label{fig6} }
\end{figure}

First, while viewing AIA movies of the emerging active region, we noticed a small eruption from the east side around 10\,UT. Figure \ref{fig5}
shows multi-wavelength AIA images of the mini-eruption. The figure is linked to a 30\,min animation in the same format.
The eruption is best seen in the 304\,\AA\, images, and in temperatures below $\sim$2\,MK (formation temperature of the 211\,\AA\, band).
Prior to the eruption the region is bipolar with a closed field cusp shaped arcade -- as was shown in Figure \ref{fig1} left panel.
After the eruption, the loop arcade appears to spread open and rapidly develop. Loops appear to draw back from the location where
the  upflows are later observed, and start to interact with the closed field to the south east: a process that seems to be integral to 
the expansion of the  upflows after April 1. Even as early as shortly after the eruption on March 31 it appears that large scale, open, 
or distantly connecting long loops develop. The field is predominantly positive in the core of the AR at this time, with only weak scattered
negative flux to the north so that field lines from the south east can only connect far from the region i.e. the negative polarities that
counterbalance the positive field are not nearby. This is the picture we also get from a magnetic field extrapolation we attempted, 
though unfortunately the AR is too close to the east limb for the model to be convincing and reliable - so we do not include it here.

The features of the small eruption can be seen in the space-time intensity plot of Figure \ref{fig6}, 
which shows high cadence (12\,s) 304\,\AA\, data obtained between 10--12\,UT\, on March 31. The plot is made by stacking intensities extracted
along the thin blue line shown in Figure \ref{fig5}. Relatively stable structures, such as the active region itself, appear as broad
horizontal trails in the plot, whereas dynamic features that move rapidly along the line appear as streaks. The sky blue arrow $P0$ points
out the streak at the start of the small eruption just after 10\,UT\, (around 60$''$ along the slice-line). The fuzzier `W' shaped streaks following $P0$ are a result of the loops
spreading open and then retracting.

\begin{figure}[h]
\centering
\includegraphics[viewport = 90 0 550 512,clip=true,width=0.50\textwidth]{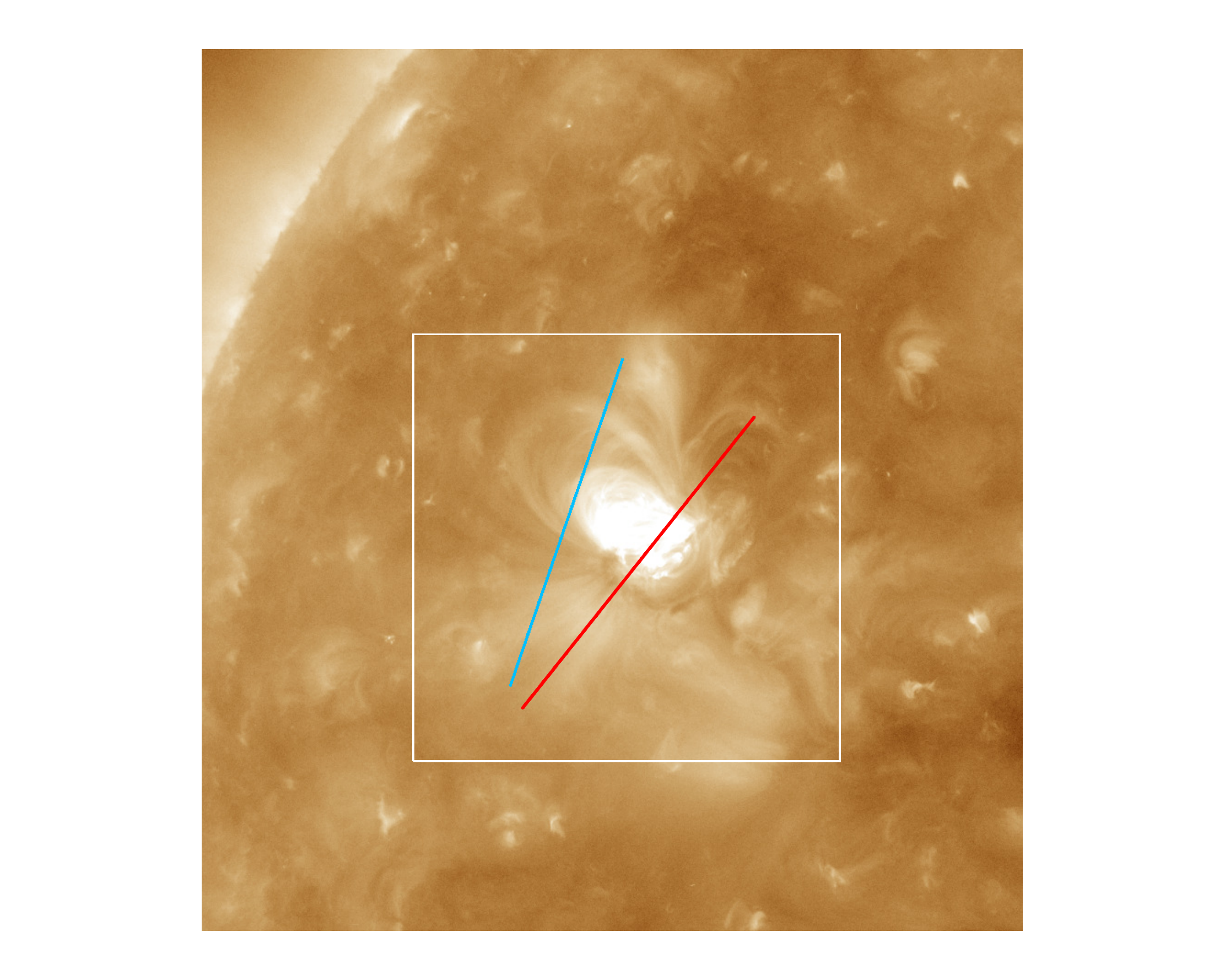}
\caption{ AIA 193\,\AA\, context for the space-time and correlation analysis, taken at 07\,UT on April 1. The 
sky blue line corresponds to the space-time plot shown in the left panel of Fig. \ref{fig8}. The red line corresponds
to the space-time plot shown in the right panel of Fig. \ref{fig8}. The white box shows the region used for the 
correlation analysis in the lower panel of Fig. \ref{fig8}. 
\label{fig7} }
\end{figure}

To quantitatively show when the active region emerged and infer when the  upflows developed, we examined space-time intensity cuts of the AIA 193\,\AA\, data.
Figure \ref{fig7} shows a context image with two slice-lines overlaid in sky blue and red. We examined several slice-lines but these two illustrate the
behavior of the eastern  upflow. Note that they are not optimised for the  upflows on the western side. The sky blue line runs across the eastern dark lane 
where EIS later detects  upflows. The idea here is to try to trace back, from when we know the dark emission is associated with  upflows detected by EIS,
to as early as we can in the active region development. The red line runs across the central loop arcade to try to trace the dark emission farther back
in time when AR 12737 was small, and the  upflows, if formed, would be closer to the core. As the lines run across the upflow and bright core we expect
that when they are stacked in time we should see a bright trail across the space-time plot that represents the core region, and a dark trail below it
that represents the area where the upflow is later detected. 
We used 193\,\AA\, data at 156\,s cadence for this analysis.

\begin{figure*}[h]
\centering
\includegraphics[viewport = 0 60 600 452,clip=true,width=0.49\textwidth]{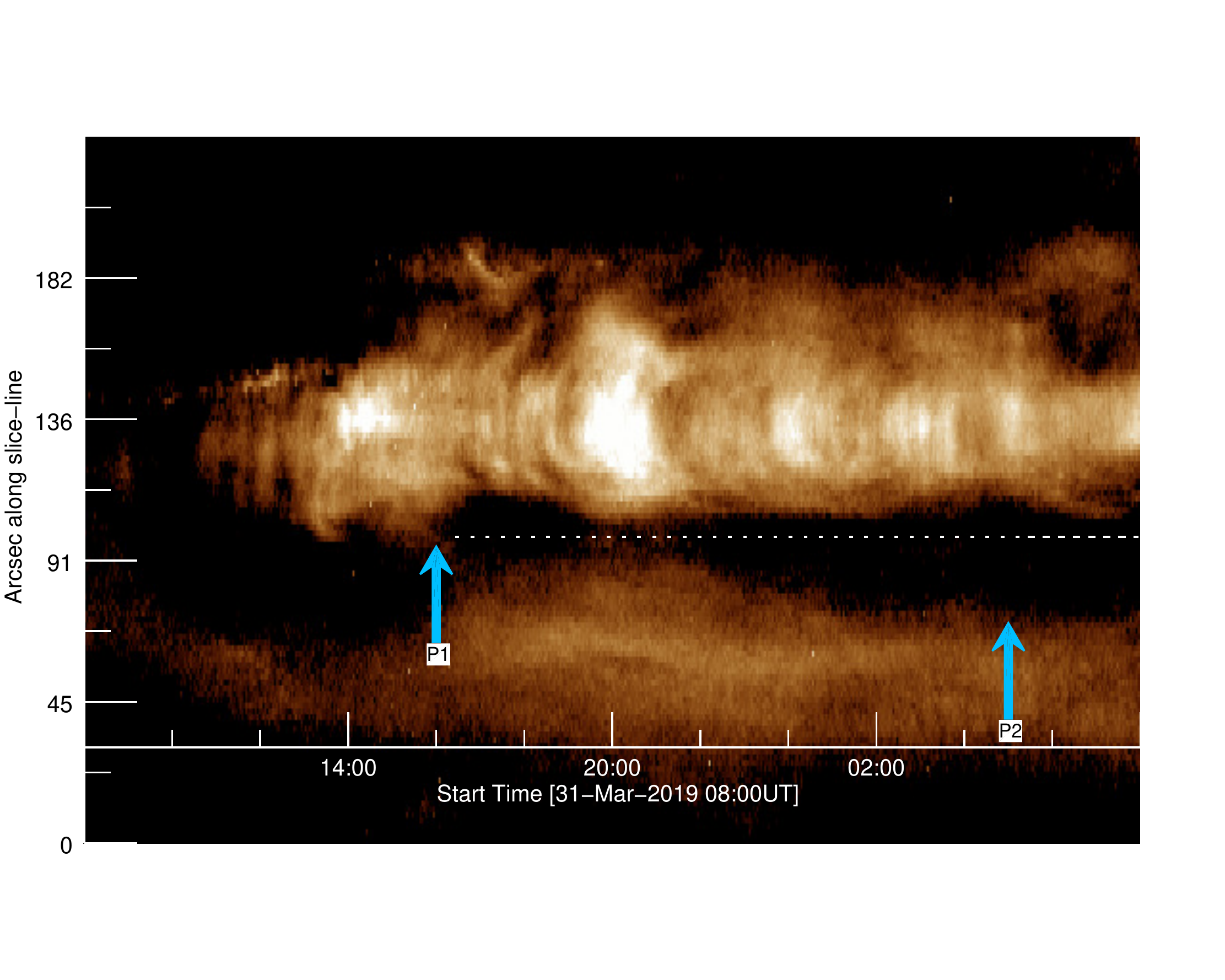}
\includegraphics[viewport = 100 15 540 482,clip=true,width=0.49\textwidth]{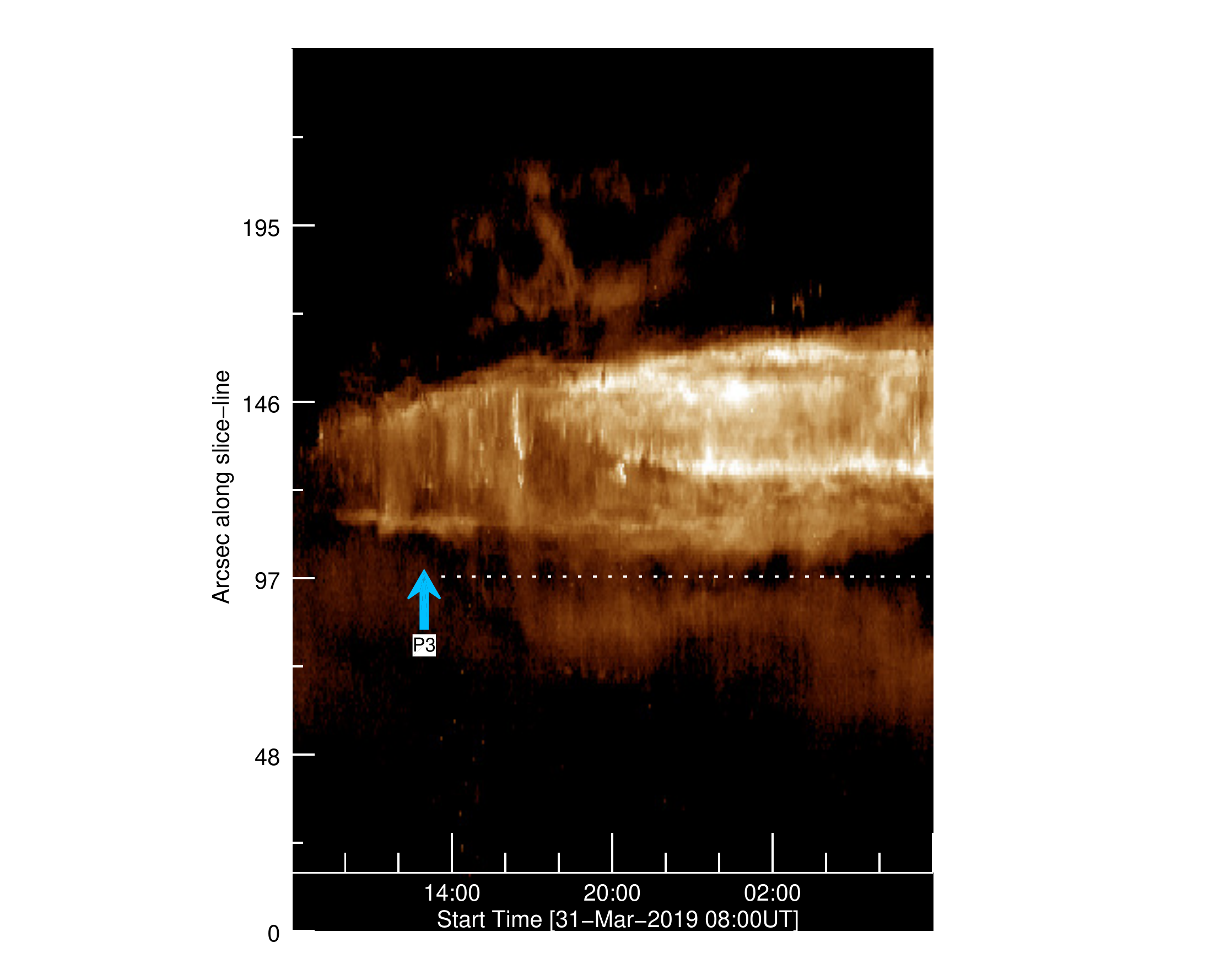}
\includegraphics[viewport = 160 0 1446 360,width=1.1\textwidth]{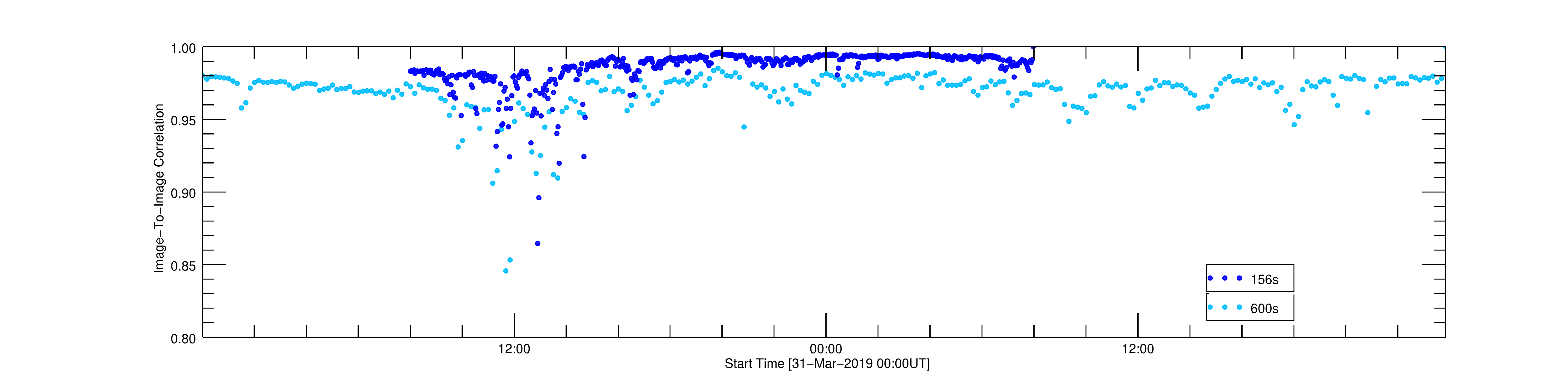}
\includegraphics[viewport = 165 0 1441 360,width=1.1\textwidth]{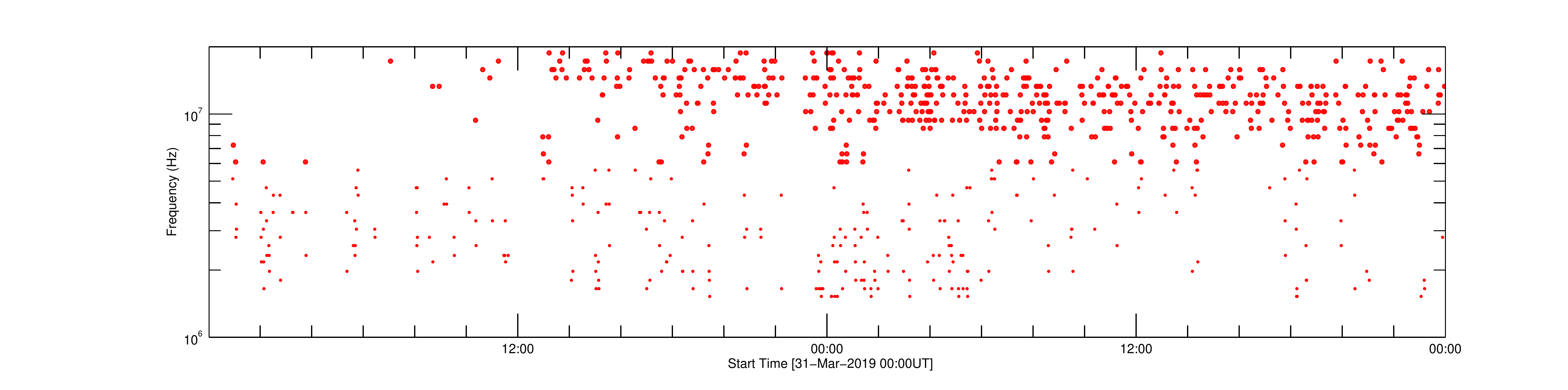}
\caption{ Space-time and correlation analysis. 
{\it Top left}: space-time plot corresponding to the sky blue line in Fig. \ref{fig7}. 
{\it Top right}: space-time plot corresponding to the red line in Fig. \ref{fig7}. 
{\it Middle panel}: image-to-image correlation at 156\,s (blue dots) and 600\,s (sky blue dots)
cadence for the region shown by the white box in Figure \ref{fig7}.
The sky blue arrows in the top panels indicate features discussed in the text.
{\it Bottom panel}: RFS data showing the onset of the type III radio storm detected by PSP. 
The plot shows the frequency of the peak normalized intensity (red) during the same time-interval as the correlation analysis.
Data points with frequency above 6$\times$10$^{6}$\,Hz are increased in size to visually highlight the noise storm.
\label{fig8} }
\end{figure*}

Figure \ref{fig8} shows the results. The top left panel is the space-time plot for the sky blue line in Figure \ref{fig7}. We can see the emergence of the
active region from $\sim$10--11\,UT and the development of the bright loop arcade $\sim$14\,UT\, in the center of the space-time plot, as expected.
The loop arcade further brightens and expands thereafter, trailing left to right across the space-time plot (approximately between the 91$''$ and 182$''$ tick marks on the Y-axis).
We point out a dark trail below the bright trail with the sky blue arrows $P1$ and $P2$ (crossing left to right and centered approximately on the 91$''$ tick mark).
This is the trace of the dark areas at the eastern edge of the active region where we expect upflows should be observed. To reiterate, this is simply because of the usual correlation between 
low intensity and blue-shifts observed previously in many active regions \citep{Doschek2008}.
For ease of identification we have also added a dotted line in the dark trail. As discussed, EIS does indeed later detect upflows in the Eastern region.
It seems that the dark region can be traced back as far as the first $P1$ arrow $\sim$16\,UT.

The top right panel shows the space-time plot for the slice-line closer in to the active region core (red in Figure \ref{fig7}).
Being closer to the location of flux emergence, this slice-line shows the appearance of the active region earlier (before 10\,UT).
Since the region is emerging at this time, any  upflows that are forming are also more likely to be obscured by the growing loop arcade. 
We can see this in the figure. The dark trail, again crossing left to right and here approximately centered on the 97$''$ Y-axis tick mark, brightens and
fades due to the line-of-sight intereference by the bright loop arcade. Nevertheless, the
sky blue arrow, $P3$, points out that the dark trail of the eastern  upflow can be traced back to around $\sim$12--13\,UT. We again added a dotted line
for ease of identification.

The middle panel of Figure \ref{fig8} shows results from our image-to-image correlation analysis. The idea here is to establish when the active region
has formed its basic structure (likely with  upflows). We compute the linear Pearson correlation coefficient, $r$, between successive images for two time
periods (08\,UT March 31--08\,UT April 1 and 00\,UT March 31--00\,UT April 2) at two different cadences (156\,s and 600\,s) inside the boxed region shown in Figure \ref{fig7}. 
When there are large changes between successive images, $r$ will decrease, and when there 
are minimal changes, $r$ will increase. We use the higher cadence data to ensure we capture any rapid changes. As we can see from the figure, the two datasets
show essentially the same result. The correlation coefficient between images is high, $r$ above 0.95 for most of the time-series, 
as we would expect given that AR 12737 was a quiescent region. Both
time series also show that flux emerged rapidly between $\sim$10--13\,UT. 
Dynamic activity during emergence caused $r$ to fall as low as $\sim$0.85, but changes began to reduce after this time and $r$ returned to its typical
value by 14--16\,UT. These timescales and periods are in line with what we discerned from the AIA movies and space-time plots. AR 12737 does not
appear to alter significantly from this time until EIS observes the  upflows.

The lower panel of Figure \ref{fig8} shows data from FIELDS/RFS for the same time-period as the image-to-image correlation analysis. The data are
derived from radio Stokes $I$ intensities detected by RFS in its two highest frequency bins (18.28--19.17\,MHz). The plot shows the frequencies of
the maximum of the mode normalized (over 40\,s windows) Stokes $I$ intensities. As discussed by \cite{Harra2021}, the radio noise storm detected 
at PSP appears most likely associated with the eastern  upflow as it expands from April 1 to April 4. The drift to lower frequencies with time also
suggests the emission height, or lateral expansion, is increasing. Looking here at the time period closer to 
active region emergence, sporadic indications of the start of the high frequency ($>$10\,MHz) radio emission are seen as early as 8--10\,UT, while the 
noise storm is clearly visible from 13\,UT\, onwards. If the radio noise storm is associated with the  upflows, then the RFS data independently support
the conclusion that the  upflows formed in the 10--13\,UT\, period as the active region developed following the small eruption. The radio emission drift
to lower frequencies is already visible on March 31 -- April 1. 

\section{Summary and Discussion} \label{sec:summary}
We have studied the formation and lifetime of the eastern  upflow from AR 12737 using EUV spectroscopic observations from EIS, NUV \& FUV spectroscopic
data from IRIS, images from AIA, and
radio data from FIELDS/RFS. Our goal was to understand where and how quickly  upflows form in an active region, and how
long they might contribute to the slow solar wind during the observed lifetime.

These observations establish the following timeline.
The active region emerged and developed from a cusp shaped loop arcade from 8\,UT\, on March 31. A small eruption occurred $\sim$10\,UT, and the typical
structure of an active region with bright core loops and peripheral dark features (usually associated with  upflows) formed from 12--16\,UT. Space-time 
intensity plots and image-to-image correlation analysis support this picture and show that the active region did not alter appreciably until EIS observed
it at 15:50\,UT\, on April 1. At this time, EIS confirmed the presence of blue-shifted upflows associated with 
large-scale field lines connecting out of the AR, from both the east and west edges, in force-free 
field models (Figure \ref{fig3}). IRIS observed the region $\sim$13.5\,hours later, and detected signatures of the  upflow in the chromosphere and transition
region. The active region grew and the  upflow area expanded between April 1--4. The magnetic field modeling is consistent with this expansion, and there is
an associated increase in the number of large-scale (potentially open) field lines \citep{Harra2021}. The eastern  upflow
was still present when the region was last observed by EIS at 06:39\,UT\, on April 9. The region was observed for 9 days.

We conclude that the  upflows in AR 12737 formed early in its lifetime (no later than 32\,hours after emergence)
and persisted for as long as EIS tracked it (85\% of the observed lifetime). 
Any contribution to the slow solar wind is therefore not a short lived phenomenon.

The lack of spectroscopic data within the first 32\,hours makes it difficult to confirm the exact time of  upflow formation, but our analysis also 
suggests that it occurred earlier. Based on the space-time and correlation analysis, the eastern  upflow can be traced back to when the typical
structure of the active region was formed between 12--16\,UT\, on March 31. That is, the  upflow may have formed as little as 4--8\,hours after
emergence and persisted for 95\% of the observed lifetime. The small eruption could have opened the magnetic field on the eastern side before this,
and the onset of the radio noise storm detected by FIELDS/RFS occurred at the start of this period. We should add the caveat that the  upflow
might not contribute to the slow wind all the time it is observed, but magnetic modeling of the region suggests it is associated with large-scale
expanding field lines the whole time that it was tracked, so in principle the plasma flows can become outflows and escape to the heliosphere.

The evidence also suggests that the  upflow formation occurs low down in the atmosphere. 
The mini-eruption ejected from the base of the cusp loop arcade soon after
emergence while the active region was still small. It was also best observed in the AIA filters associated with cooler temperatures, especially 
304\,\AA. Even if we only consider the spectroscopic data,
possible signatures of the  upflows were observed in the chromosphere and transition region by IRIS on April 2, and the EIS data show that 
the active region and  upflow did not grow to their full extent until April 4. This implies that the  upflow formation was well underway before AR 12737
had expanded to interact with high lying magnetic fields. The radio noise storm observed by FIELDS/RFS also showed a frequency drift that can be
interpreted as the emission height forming at lower altitudes.
Future multi-mission observations, of the earliest stages of active region emergence, will hopefully pin down more accurately some of the suggestions put forward in this article.


\acknowledgments 
This study benefited from discussions at the meeting of ISSI international team 463; project title `Exploring The Solar Wind In Regions Closer Than Ever Observed Before'.
The work of DHB and HPW was funded by the NASA Hinode program. VP acknowledges support by NASA contract NNG09FA40C (IRIS) and grant no. 80NSSC21K0623. CM acknowledges
financial support from the Argentine grants PICT 2016-0221 (ANPCyT) and
UBACyT 20020170100611BA (UBA). This work is supported by the Swiss National Science Foundation - SNF. 
Hinode is a Japanese mission developed and launched by ISAS/JAXA, with NAOJ as domestic partner and NASA and STFC (UK) as international partners. 
It is operated by these agencies in co-operation with ESA and NSC (Norway). The SDO data are courtesy of NASA/SDO and the AIA, EVE, and HMI science teams.
IRIS is a NASA small explorer mission developed and operated by LMSAL with mission operations executed at NASA Ames Research center and major contributions to downlink 
communications funded by ESA and the Norwegian Space Centre. 
We acknowledge the NASA Parker Solar Probe Mission and FIELDS team led by S.D.Bale for use of data.
The PSP/FIELDS experiment was developed and is operated under NASA contract NNN06AA01C.


\end{document}